\documentclass[
  ,draft            
  ]
  {aipproc}

\layoutstyle{6x9}

\def\beq{\begin{equation}}
\def\eeq{\end{equation}}
\def\beqn{\begin{eqnarray}}
\def\eeqn{\end{eqnarray}}

\begin{document}

\title{Antigravitation}


\author{Sabine Hossenfelder}{
  address={Nordita, Roslagstullsbacken 23, 106 91 Stockholm, Sweden}
}

\begin{abstract}
We discuss why there are no negative gravitational sources in General Relativity and show
that it is possible to extend the classical theory such that repulsive gravitational
interaction occurs. This is the summary of a talk given at the 17th International Conference on Supersymmetry 
and the Unification of Fundamental Interactions in Boston, June 2009. 
\end{abstract}

\maketitle


\section{Why no negative gravitational charges?}

In General Relativity (GR) the gravitational charge of the source, its energy, is always positive. If one
considers the Newtonian potential describing the interaction between two point masses, it seems straightforward
to introduce negative gravitational masses: to obtain a 
repulsive interaction, just change the sign of one of the masses. Unlike Electrodynamics which is mediated by a spin-1 field, the gravitational interaction is transmitted by a spin-2 field, thus like charges will attract and unlike charges repel.

So far, so good. But if one takes into account the motion of a testparticle in a gravitational background field, one knows that this curve, a geodesic, is independent of the mass of the particle. If the curve does not depend
on the particle's mass then a particle of negative gravitational mass, henceforth called `antigravitating,' would fall
towards a gravitating source of either kind. One then had a situation in which a gravitating and an antigravitating particle did both, attract and repel \cite{Bondi}; a scenario that is very implausible.

Another common objection to antigravitation is that the existence of negative 
masses would allow for infinite production of
particle pairs. Because energy conservation does not forbid production of a zero net sum out of nothing,
a disastrous vacuum decay would result. While this is a quantum effect and the following discusses
a merely classical scenario, it is a serious concern that will be briefly addressed later. 

A frequently occurring misunderstanding stems from the definition of antigravitation, so some words of
clarification are in order. Antigravitation has sometimes been used \cite{Feynman} to mean a repulsive gravitational
interaction between particles and their antiparticles. While experimental data on the gravitational
interaction of antiparticles is not entirely conclusive, a different gravitational behavior of antiparticles is 
from a quantum field theoretical perspective not very plausible. What is here referred to as antigravitating particles instead is an entirely new type of matter that, in the simplest case, would be a copy of the Standard Model (SM). These particles are identical to the particles of the SM except for their gravitational interaction. They attract
each other but repel our usual matter. 

The gravitating and antigravitating
particles interact only gravitationally. Thus, not only would the antigravitating matter be dark, the 
interaction is also too weak for these particles to be produced in experiments on Earth. Such matter would
be detectable only by its gravitational effects. Since it was repelled by our
normal matter during the formation of structures in the universe, one would not expect it to
be present in our vicinity in sizeable amounts. It would instead collect in what we think are
voids.

We will in the following propose such an extension of GR with negative gravitational charges based on the
assumption that the antigravitating particles are just like the common particles, and that exchanging
both sorts of particles would not alter the physics. We are thus postulating a symmetry between
gravitating and antigravitating particles. This does not necessarily mean though that the amount of
both types of matter in our universe is the same. 

Since we know from cosmological and astrophysical
observations that our understanding of the matter content of the universe and/or its gravitational
dynamics is incomplete, such a symmetry is appealing for both theoretical as well as experimental
reasons. It is thus not surprising it has been studied before in various approaches \cite{linde,Petit:1995ys,Drummond:2001rj,Moffat:2002kj,Quiros:2004ge,HenryCouannier:2006qn}. The scenario
outlined in the following, and explained in more detail in \cite{Hossenfelder:2008bg}, has the virtue of being entirely covariant and not allowing vacuum decay because it does not necessitate negative kinetic energies.

\section{How to add negative gravitational charges}
 
As discussed above, an antigravitating particle that moves on a geodesic is implausible. However, a geodesic
is uniquely defined only through the connection employed, and the connection itself is uniquely
defined only by requiring it to be torsion-free and metric-compatible. Since torsion does not
affect geodesics, one wants a second connection that is torsion-free but not metric-compatible. To that end, 
introduce a second metric ${\bf \underline h}$ over the manifold, and
construct a second derivative, denoted ${}^{\underline h}\underline \nabla$, that is compatible, not with the usual metric $\bf g$, but with the
second metric ${\bf \underline h}$. The antigravitating particles will feel a background with distance measures defined by the
second metric and move on geodesics defined according to this metric. With this derivative one
can then define a second curvature tensor and its contractions as usual. 

We refer in the following to the antigravitating (normally gravitating) fields, particles and observers as $h$-fields, $h$-particles, $h$-observers etc ($g$-fields, $g$-particles, $g$-observers). It is straightforward to write down a Lagrangian for the
$h$-fields by replacing the usual metric with the second metric and the usual derivative with the corresponding second derivative. One should not forget to also replace the volume-element, since otherwise Gauss's law cannot be applied
and the equations of motion one obtains are not symmetric to the usual ones. For a massless scalar $h$-field $\underline \phi$ the action then looks like 
\beqn
S = \int d^4  x \sqrt{- {\underline h}}~  h^{{\underline{\nu \kappa}}} ~ {}^{(\underline h)}\underline \nabla_{\underline{\kappa}} \underline \phi {}^{(\underline h)}\underline \nabla_{\underline{\nu}} \underline \phi \quad .
\eeqn
These $h$-fields are tensors over the manifold and can be expanded in
a local basis of the tangential space. However, there is no tool to contract the basis of the tangential
space in which the $h$-fields are expanded with those of the $g$-fields. The local bases
of both spaces are trivially isomorphic in each point, but they cannot be directly compared to tell whether they are
indeed the same. This is conceptually similar to Special Relativity. We have two observers. Here, one is
constituted or normal matter, the other one of antigravitating matter. They do both describe their
physics in tensor equations and we know these equations have the same form for both observers. Yet to
compare them, we need to find a transformation from one set of observables to the other. These 
transformations to pull over the tensors of the antigravitating observer to that of the normally gravitating, and vice versa we will
call the `pull-overs.'  

The pull-overs do assign observables for $h$-fields to the $g$-observer, and observables
for $g$-fields to the $h$-observer. By this, they do preserve the tensor structure of objects, and also the
covariance of derivatives. In particular, for the $g$-observer they assign a 2-tensor ${\bf h} = P_{\underline h}({\bf \underline h})$ to the second metric ${\bf \underline h}$, and similarly
the $h$-observer assigns a pulled-over 2-tensor ${\bf \underline g} = P_g({\bf g})$ to our metric ${\bf g}$. Note underlines. The
underlined quantities are the observables of the $h$-observer, but cannot be observed directly by the $g$-observer. Only
after applying the pull-over are they converted into standard tensor-fields. The situation is the same the other
way round. 

We further define a map ${\bf a}$ that transforms the one metric into the pull-over of the other  
\beqn
g_{\epsilon \lambda}  = a_{\epsilon}^{\;\;\nu} a_{\lambda}^{\;\;\kappa} ~ h_{\nu \kappa}   \label{haga} \quad.
\eeqn
Since both ${\bf g}$ and ${\bf h}$ are symmetric, ${\bf a}$ is not completely determined by (\ref{haga}). We
fix the remaining six degrees of freedom by requiring it to be symmetric, i.e. $g^{\kappa \nu} a^{\epsilon}_{\;\;\nu} = a^{\epsilon \kappa} = a^{\kappa \epsilon}$. We can pull over ${\bf a}$ by
\beqn
a_{\underline \epsilon}^{\;\;\underline \nu} = 
\left[P_{g}\right]^{\epsilon}_{\;\;\underline \epsilon} a_{\epsilon}^{\;\; \nu}\left[ P_{\underline h} \right]^{\underline \nu}_{\;\;\nu}  \quad,
\eeqn
which then gives the relation
\beqn
g_{\underline{\epsilon \lambda}} = a_{\underline \epsilon}^{\;\; \underline \nu} a_{\underline \lambda}^{\;\; \underline \kappa} ~ h_{\underline{\nu \kappa}} \label{hagau}  \quad.
\eeqn
This pulled-over quantity is also required to be symmetric. It is further useful to define a combination of ${\bf a}$ and the pull-overs that maps ${\bf g}$ to ${\bf \underline h}$ via
\beqn
a_{\epsilon}^{\;\;\underline \nu}  = a_{\epsilon}^{\;\;\nu} [P_{\underline h}]^{\underline \nu}_{\;\;\nu} \quad, \quad  g_{\epsilon \lambda} = a_{\epsilon}^{\;\;\underline \nu} a_{\lambda}^{\;\;\underline \kappa} ~ h_{\underline{\nu\kappa}}  \quad. \label{gauha}
\eeqn
And by raising and lowering some indices we also have
\beqn
g_{\epsilon \lambda} = a_{\epsilon}^{\;\;\underline \nu} a_{\lambda \underline \nu}~,~ 
 h_{{\underline{\nu \kappa}}} = a^{\epsilon}_{\;\;\underline{\nu}} a_{\epsilon \underline{\kappa}}  
  \quad.
\eeqn
This introduced map ${\bf a}$ is a convenience and not an independent dynamical field, since it is defined 
by relating ${\bf g}$ to $P_{\underline h}({\bf \underline h})$. 

Now that the basics of the bi-metric formalism is in place, the field equations need to be found to determine
the equations of motion for the second metric. We thus extend Einstein's field equations by adding a negative 
source term constituted of the antigravitating type of matter. Then, making use of the symmetry
principle, we add a second set of equations for the second metric. For this metric, it is the $h$-source
that has the usual sign, whereas our $g$-fields have the negative sign. 

However, if one simply does this, the system of equations is over-constrained due to the contracted Bianchi identities. 
These two times four equations require the additional source terms to be covariantly conserved, yet there are no more degrees of freedom left. The reason for this inconsistency is the neglect to pull over the source terms which adds additional degrees of freedom. Together with the requirement of torsion-free-ness the pull-overs provide 
the missing two times four degrees of freedom. In special cases, the pull-overs are just the identity but in
general they are non-trivial. The field equations then read
\begin{eqnarray}
{}^{(g)}R_{\kappa \nu} - \frac{1}{2} g_{\kappa \nu} {}^{(g)}R &=& 8 \pi G \left( T_{\kappa \nu} - \sqrt{\frac{h}{g}} a_\nu^{\;\;\underline \nu} a_\kappa^{\;\;\underline \kappa} \underline T_{\underline{\nu \kappa}} \right) \label{fe1} \\
{}^{(h)}R_{\underline{\nu \kappa}} - \frac{1}{2} h_{\underline {\nu \kappa}} {}^{({\underline h})}R &=& 8 \pi G \left( \underline T_{\underline{\nu \kappa}} - \sqrt{\frac{\underline g}{{\underline{h}}}} a^{\kappa}_{\;\underline{\kappa}} a^{\nu}_{\;\underline{\nu}} T_{\kappa \nu} \right) \quad, \label{fe2}
\end{eqnarray}
with 
\begin{eqnarray}
T_{\mu \nu} &=& - \frac{1}{\sqrt{-g}} \frac{\delta {\cal L}}{\delta g^{\mu\nu}} + \frac{1}{2} g_{\mu \nu} {\cal L} \quad,\quad
\underline T_{\underline{\nu \kappa}} = - \frac{1}{\sqrt{-\underline{h}}} 
\frac{\delta \underline {\cal L}}{\delta 
h^{\underline{\nu \kappa}}} + \frac{1}{2} h_{\underline{\nu \kappa}} \underline {\cal L} \quad. \label{set2}
\end{eqnarray}
Also note the factors converting the measures of the stress-energy tensors in Eqs. (\ref{fe1},\ref{fe2}), since they are densities.

There is an action principle from which the above equation can be derived \cite{Hossenfelder:2008bg}. 
Though the constrained space of this summary is insufficient for more details, it should be mentioned that the 
action does not have any negative kinetic energy terms. The minus sign in the source term stems from the 
variation over the both metrics, together with the requirement that the variation of the ${\bf a}$'s vanishes, i.e. $\delta a^{\nu\kappa} = \delta a^{{\underline{\nu\kappa}}} = 0$. 
The change of sign thus appears only for the sources of the gravitational field. It does not appear if one 
takes the variation with respect to the fields which provides the stress-energy that can be considered a 
generalization of `inertial' mass. Then, to finally come back to the concern about an unstable vacuum, both the gravitational and the inertial mass of particles is conserved separately which means in particular it is not 
possible to produce a pair of gravitating and anti-gravitating particles out of the vacuum. 

{\bf Note added:} A no-go theorem for bi-metric gravity with positive and negative mass was recently
put forward in \cite{Hohmann:2009bi}. It has
been shown in \cite{Hossenfelder:2009hb}, that this does not affect the here discussed model.


\bibliographystyle{aipproc}

\end{document}